\documentstyle[11pt,newpasp,twoside,epsf]{article}
\def\edcomment#1{\iffalse\marginpar{\raggedright\sl#1\/}\else\relax\fi}
\reversemarginpar

\newcommand{\nswd}{NS$-$WD}
\newcommand{\rb}{${\cal R}_{\rm b}$}

\begin{document}
\title{The Galactic Formation Rate of Eccentric Neutron Star -- 
White Dwarf Binaries}
 \author{V. Kalogera$^1$, C. Kim$^1$, D. R. Lorimer$^2$, M. Ihm$^1$, K. Belczynski$^{1,3}$}
\affil{$^1$Northwestern University, Department of Physics and Astronomy, 2145 Sheridan Rd. Evanston, IL, 60201, USA}
\affil{$^2$University of Manchester, Jodrell Bank Observatory, Macclesfield,
Cheshire, SK11 9DL, UK}
\affil{$^{3}$Northwestern Lindheimer Fellow}

 \begin{abstract}
In this paper we consider the population of eccentric binaries with a
neutron star and a white dwarf that has been revealed in our galaxy
in recent years through binary pulsar observations. We apply our
statistical analysis method (Kim, Kalogera, \& Lorimer 2003) 
and calculate the Galactic formation rate of these binaries
empirically. We then compare our results with rate predictions based
on binary population synthesis from various research groups and for
various ranges of model input parameters.
For our reference moel, we find the Galactic formation rate of these
eccentric systems to be $\sim$ 7~Myr$^{-1}$, about an order of
magnitude smaller than results from binary evolution
estimations. However, the empirical estimates are calculated with no
correction for pulsar beaming, and therefore they should be taken as
lower limits. Despite uncertainties that exceed an order of magnitude,
there is significant overlap of the various rate calculations. This
consistency lends confidence that our current understanding of the
formation of these eccentric \nswd~binaries is reasonable.
\end{abstract}

\section{Introduction}

Binary pulsars with white dwarf companions (\nswd~binaries) in {\it
eccentric} orbits have been revealed with binary pulsar and optical
observations in recent years.  This sub-population is considered to be
rather special because, based on our standard understanding of binary
evolution, \nswd~binaries are expected to be circular: a neutron star
forms first from the original binary primary and the white dwarf
formation follows mass transfer episodes that are expected to
circularize the binary orbit. For a review of this scenario,
see Lorimer (2001). The neutron star produced by this evolutionary path is
expected to be `recycled', i.e.~spun-up by mass accretion from its
companion.  However, the existence of eccentric \nswd~binaries such as
PSR~B2303+46 (Stokes, Taylor, \& Dewey 1985; van Kerkwijk \& Kulkarni
1999) and PSR~J1141--6545 (Kaspi et al.~2000; Bailes et al.~2003) implies
that a different evolutionary path from the standard scenario is also
be possible (Tutukov \& Yungelson 1993; Portgies Zwart \& Yungelson
1999; Tauris \& Sennels 2000; Nelemans, Yungelson, \& Portegies Zwart
2001; Brown et al.~2001; Davies, Ritter, \& King 2002). The non-zero
eccentricity of the binary orbit is introduced by a supernova (SN)
explosion of the secondary companion that occurs after the original
primary has already evolved into a white dwarf.

Our motivation for this study is to estimate the Galactic formation
rate of eccentric binaries based on the observed pulsars. We apply our
statistical analysis developed to estimate the Galactic
double-neutron-star (DNS) merger rate (Kim, Kalogera, \& Lorimer 2003)
and derive a probability density function (PDF) of the Galactic formation 
rate for eccentric \nswd~binaries. This method provides us with rate
estimates that are independent from those obtained using binary
evolution calculations.  We compare our empirical rate estimates to
results from population synthesis calculations and conclude that,
despite the large uncertainties, the results are indeed consistent.

\section{Formation of eccentric \nswd~binaries} 

In our classical understanding of binary evolution, we expect that the
formation process of \nswd~binaries in close orbits involves the
circularization of the binary orbit, even for systems with massive
white dwarfs: the neutron star forms first and possibly induces an
eccentricity, but subsequent mass transfer from the white-dwarf
progenitor is expected to circularize the orbit (Verbunt \& Phinney
1995). Indeed PSR~J0751+1807 (Lundgren, Zepka, \& Cordes 1995; Nice,
Splaver, \& Stairs 2004) and PSR~J1757$-$5322 (Edwards \& Bailes
2001) have very small eccentricities ($e \le 10^{-4}$). However, the
discovery of the white dwarf companion to PSR~B2303+46 (van Kerkwijk
\& Kulkarni 1999) led various groups to consider a different
evolutionary path that could explain the observed eccentricity of
$e=0.658$ (Portgies Zwart \& Yungelson 1999; Tauris \& Sennels 2000;
Nelemans, Yungelson, \& Portegies Zwart 2001; Brown et al.\ 2001;
Davies, Ritter, \& King 2002). 

In these modified scenarios,
the original primary star, which is
massive enough to form a white dwarf but not a neutron star, evolves
first and transfers its mass to the secondary star. After the primary
star loses its envelope it becomes a white dwarf, while the secondary
continues its evolution but at a higher mass than its initial mass
(high enough to form a neutron star) due to the mass transfer phase
from the original primary. The primary eventually fills its Roche lobe and a
common envelope phase ensues: the white dwarf and the helium core of
the secondary spiral together as the common envelope of the two stars
is ejected from the system. The outcome is a tight binary with a white
dwarf and a helium star that is massive enough to explode in a
supernova. This introduces an eccentricity to the orbit of the
\nswd~binary that forms. Hence eccentric \nswd~binaries are formed
because of two main facts relevant to the early evolution of the
binary progenitor: (1) the initial binary components are both massive
enough to form only a white dwarf, but still close to being massive
enough to form a neutron star; (2) the initial mass transfer phase
from the primary to the secondary increases the mass of the secondary
enough that it can now form a neutron star eventually, but only after
the initial primary ends its evolution as a white dwarf.

\begin{table}[t]\small
  \begin{center}
    \caption{\footnotesize Observational properties of eccentric
    NS--WD binaries. The columns indicate the
    pulsar name, spin period $P$, spin-down rate $\dot{P}$, orbital
    period $P_b$, the estimated mass of the WD companion $m_c$,
    orbital eccentricity $e$, characteristic age $\tau_c$, time to
    reach the death line $\tau_d$.
    }

    \vskip20pt
    \begin{tabular}{lcccccccr}
    \tableline
     PSRs & $P$ & $\dot{P}$ & $P_{\rm b}$ & m$_{\rm c}$ & $e$ & $\tau_{\rm c}$ & $\tau_{\rm d}$ \\
     \tableline
      {} &  (ms) &  (s~s$^{-1}$) &  (hr) & (M$_{\rm \odot}$) & {} & (Myr) & (Gyr) \\
     \tableline
      J1141$-$6545 & 393.9 & 4.29$\times 10^{-15}$ & 4.744 & 0.986 & 0.172 & 1.45 & 0.104 \\
      B2303+46     & 1066  & 5.69$\times 10^{-16}$ & 296.2 & 1.24  & 0.658 & 29.7 & 0.140 \\
     \tableline
     \end{tabular}
   \end{center}
\end{table}

\section{Empirical \nswd~formation rate estimates}

In KKL, we introduced a statistical method to calculate a PDF of the
rate estimates for Galactic close DNS systems.  This method can be
applied to any type of pulsar population of interest.

Here, we consider eccentric \nswd~binaries. In order to calculate
their Galactic formation rate \rb, we need to estimate: (1) the number
$N_{\rm tot}$ of Galactic pulsar populations with pulse and orbital
characteristics {\it similar\/} to those in the observed sample of
eccentric \nswd~(i.e.\ J1141--6545 or B2303+46); (2) the lifetime
$\tau_{\rm life}$ of each observed system; (3) an upward
correction factor $f_{\rm b}$ for pulsar beaming.

We calculate $N_{\rm tot}$ by modeling in detail the pulsar survey
selection effects associated with the discovery of these systems for a
number of parent pulsar population models described in KKL. The model
assumptions for the pulsar luminosity function dominate the systematic
uncertainties of our overall calculation.

For our reference model (model 6 in KKL), we obtain the most likely
value of $N_{\rm tot}$ for PSRs J1141-6545 and B2303+46 to be $N_{\rm
1141}\simeq$370 and $N_{\rm 2303}\simeq$240, respectively.

\begin{figure}
\plotfiddle{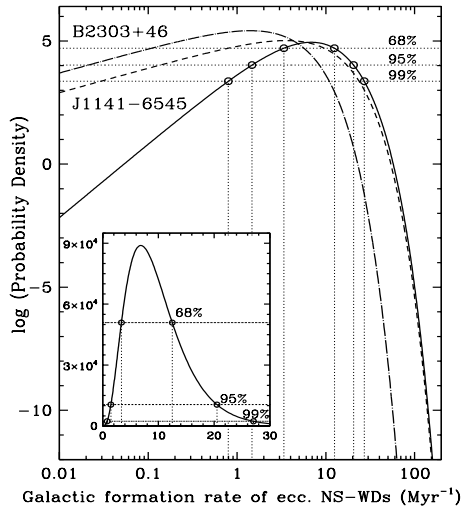}{5.9cm}{0}{30}{27}{-100}{-15}
\caption{The PDF of Galactic formation rate estimates for eccentric
\nswd~binaries (solid line) for our reference model. Dashed and
dot-dashed lines represent the individual PDFs of the formation rates
for sub-population of binaries similar to either PSR B2303+46 or
J1141$-$6545. No corrections for pulsar beaming have been applied.}
\end{figure}

The lifetime of the system is defined as $\tau_{\rm life} \equiv
\tau_{\rm c} + \tau_{\rm d}$, where $\tau_{\rm c}$ is the
characteristic age. $\tau_{\rm d}$ is the time which a pulsar will
reach the ``death line'' (Ruderman \& Sutherland 1975). We calculated
$\tau_{\rm d}$ following (Chen \& Ruderman 1993; eq.~(9) in their
paper). The calculated PDF of the formation rate is dominated by PSR
J1141-6545-like population due to both the shorter lifetime and
number abundance of this population (Fig.\ 1). The estimated formation rate
for our reference model is
\rb$=6.8^{+5.6}_{-3.5}$$^{+13.7}_{-5.6}\,$Myr$^{-1}$ at 68\% and 95\%
confidence limits, respectively. The most likely values of \rb~lie in
the range $0.5-16\,$Myr$^{-1}$ for all the PSR population models we
consider (see Table 2 for the rate estimates for a number of
models with different pulsar luminosity functions). In the absence of
observational constraints on the geometry of both pulsars, we decided
to not include any such upward correction for
pulsar-beaming. Therefore, our estimated rates should be considered as
lower limits. We can also compare these estimates with those for
the Galactic DNS rate. The most likely values of DNS rates are found 
in the range $4-224$ Myr$^{-1}$ for the models considered with pulsar beaming correction. 
(see contribution by Kim et al.\ in these proceedings and Kalogera et al. (2004)). Since the beaming correction factor for young pulsars is presumably larger than those for recycled pulsars, the true Galactic formation rate for eccentric \nswd~binaries are expected to be comparable to that of DNS systems we estimated.

Population synthesis calculations have been widely used to study the
formation of various types of compact object binaries including
eccentric \nswd~binaries. The range of details in the evolutionary
calculations as well as the extent (if any) of the parameter studies
vary significantly. In this section we summarize the main results from
theoretical studies in literature and compare them to the empirical
rates derived in the previous section.

Portegies Zwart \& Yungelson (1999) obtained a formation rate for
eccentric \nswd~binaries comparable, but somewhat larger than that of
DNS systems (\rb$=$44 and 34 Myr$^{-1}$ for eccentric \nswd~and DNS
systems respectively). Tauris \& Sennels (2000) presented similar
results (\rb$\sim57$ Myr$^{-1}$) and concluded that the formation rate of eccentric
\nswd~binaries is $\sim$18 times higher than that of DNS
systems. Davies, Ritter, \& King (2002) estimated the Galactic
formation rate of systems like J1141$-$6545 or B2303+46. They
obtained formation rates in the wide range of $\sim10^{-5}-10^{-3}$
yr$^{-1}$ when considering both J1141$-$6545$-$like and B2303+46$-$like systems
with different evolutionary histories. Finally, Nelemans, Yungelson, \& Portegies Zwart (2001) derived 240 Myr$^{-1}$ for the formation rate of all types of \nswd~binaries, which is comparable
to other studies. We consider this rate to be an upper limit of
eccentric \nswd~ (shown as an open square with a downward arrow
in Fig.~2).

In addition to the above results from then literature, we have performed
population synthesis calculations using the {\tt StarTrack} population
synthesis code (Belczynski, Kalogera, \& Bulik 2002; Belczynski et al.\ 2004). We
have explored a selection of models (more than any of the other
studies) that promise to give us the widest variations of rate
estimates. We derive a range $\sim30-100$ Myr$^{-1}$
(using an absolute
normalization of models based on empirical supernova rate estimates
for our Galaxy; Cappellaro, Evans, \& Turatto 1999).

\begin{table}[t]\small
  \begin{center}
    \caption{\footnotesize The estimated Glactic formation rate and the most likely value of $N_{\rm tot}$ of eccentric NS--WD binaries for models with different pulsar luminosity functions. Model number is same with KKL. We show the most likely value of $R_{\rm b}$ at 68\% and 95\% confidence limits. 
    }
    \vskip20pt
    \begin{tabular}{lrllrr}
    \tableline
     Model &$R_{\rm b}$ & (Myr$^{-1}$) & {} & $N_{\rm 1141}$ & $N_{\rm 2303}$\\
    \tableline
      {} & peak& 68\% & 95\% & {} & {}\\
     \tableline
      1  & 2.1  & $^{+1.7}_{-1.0}$  & $^{+4.0}_{-1.6}$  & 110 &  70\\
      {\bf 6}  & 6.8  & $^{+5.6}_{-3.5}$  & $^{+13.7}_{-5.4}$ & 370 & 240\\
      9  & 0.7  & $^{+0.6}_{-0.4}$  & $^{+1.4}_{-0.6}$  &  40 &  30\\
      10 & 2.6  & $^{+2.1}_{-1.3}$  & $^{+5.1}_{-2.0}$  & 140 &  90\\
      12 & 1.0  & $^{+0.8}_{-0.5}$  & $^{+1.9}_{-0.8}$  &  50 &  40\\
      14 & 0.5  & $^{+0.4}_{-0.2}$  & $^{+0.8}_{-0.4}$  &  20 &  20\\
      15 & 15.7 & $^{+13.1}_{-8.1}$ & $^{+31.9}_{-12.4}$& 870 & 530\\
      17 & 3.9  & $^{+3.3}_{-2.0}$  & $^{+8.1}_{-3.1}$  & 220 & 130\\
      19 & 1.1  & $^{+0.9}_{-0.6}$  & $^{+2.3}_{-0.9}$  &  60 &  40\\
      20 & 8.3  & $^{+7.2}_{-4.3}$  & $^{+17.7}_{-6.6}$ & 490 & 260\\
     \tableline
     \end{tabular}
   \end{center}
\end{table}

The different results from various population studies are mainly
attributed to varying assumptions about the initial mass function, the
initial mass-transfer process, the assumed star-formation or supernova
rate used as a normalization factor, the initial binary fraction, and
NS kicks. We also note that the lifetime of the system is a free
parameter in the theoretical rate estimation, which is attributed to
at least an order of magnitude of the uncertainty in the calculation.

\begin{figure}
\plotfiddle{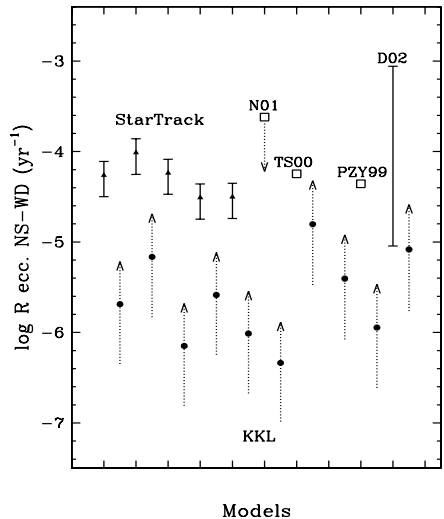}{5.9cm}{0}{30}{27}{-100}{-15}
\caption{Comparison between the empirical and theoretical rate
estimates. Error bars with filled triangles indicate results from {\tt StarTrack}, open squares and a solid line are adapted from the literature, and filled circles with error bars are obtained in this work (see text for details).}
\label{fig:birth}
\end{figure}
\section{Comparison with rates from binary evolution}
In Fig.~2, we overlap the empirical formation rate estimates for
eccentric \nswd~ binaries (filled circles with error bars labed by KKL) with
previous studied (open squares and a think solid line labeled by N01 (Nelemans, Yungelson, \& Portegies Zwart 2001), TS00 (Tauris, \& Sennels 2000), and PZY99 (Portgies Zwart \& Yungelson 1999), and D02 (Davies, Ritter, \& King 2002), respectively) as well as results from {\tt StarTrack} (thin solid lines with filled triangles).
It is encouraging
that the overall rate estimates from different methods appear to be
consistent with one another. If we consider the most likely values of
the empirical rate estimates (filled circles), they are somewhat
smaller than the theoretical values. However, at a 95\% confidence
interval, the empirical and theoretical estimates become
comparable. We also emphasize that the {\em upward} correction for
pulsar-beaming has not been applied, and therefore the empirical rate
estimates should be considered as lower limits.

In conclusion, we find that our empirical estimates for the Galactic
formation rate of eccentric \nswd~binaries are overall consistent with
the estimates derived based on binary population synthesis
models. Despite the large uncertainties we consider this consistency
as evidence that our current theoretical understanding for the
formation of eccentric \nswd~is reasonable. However, the extent of
the range covered by the empirical estimates and the population
synthesis studies that attempt even a minimal parameter study (e.g.,
Davies, Ritter, \& King (2002) and our results from {\tt StarTrack})
indicates once again the necessity for careful parameter studies of
rate calculations. It also indicates that the empirical rates could in
principle be used to constrain binary evolution calculations.

\acknowledgments
We thank the Aspen Center for Physics for its hospitality.  This
research is partially supported by NSF grant 0121420, and a Packard
Foundation Fellowship in Science and Engineering to VK. DRL is a
University Research Fellow supported by the Royal Society. He also
thanks the Theoretical Astrophysics Vistors' fund for support. 
KB is a Lindheimer Fellow at Northwestern University and 
also acknowledges support from grant PBZ-KBN-054/p03/2001.

\end{document}